\documentclass[twocolumn]{pasj00}
\draft
\usepackage[dvips]{color}

\begin{document}
\SetRunningHead{N. Narita et al.}{The Rossiter-McLaughlin Effect of XO-4b}
\Received{2010/08/23}
\Accepted{}

\title{The Rossiter-McLaughlin Effect of
the Transiting Exoplanet XO-4b$^*$}

\author{
Norio \textsc{Narita},\altaffilmark{1}
Teruyuki \textsc{Hirano},\altaffilmark{2}
Roberto \textsc{Sanchis-Ojeda},\altaffilmark{3}
Joshua N.\ \textsc{Winn},\altaffilmark{3}\\
Matthew J. \textsc{Holman},\altaffilmark{4}
Bun'ei \textsc{Sato},\altaffilmark{5}
Wako \textsc{Aoki},\altaffilmark{1} and
Motohide \textsc{Tamura}\altaffilmark{1}
}

\altaffiltext{1}{
National Astronomical Observatory of Japan, 2-21-1 Osawa,
Mitaka, Tokyo, 181-8588, Japan
}
\email{norio.narita@nao.ac.jp}

\altaffiltext{2}{
Department of Physics, The University of Tokyo, Tokyo, 113-0033, Japan
}

\altaffiltext{3}{
Department of Physics, and Kavli Institute for Astrophysics
and Space Research,\\
Massachusetts Institute of Technology, Cambridge, MA 02139, USA
}

\altaffiltext{4}{
Harvard-Smithsonian Center for Astrophysics,
60 Garden Street, Cambridge, MA 02138, USA
}

\altaffiltext{5}{
Global Edge Institute, Tokyo Institute of Technology,
2-12-1 Ookayama, Meguro, Tokyo, 152-8550, Japan
}


\maketitle

\begin{abstract}
We report photometric and radial velocity observations
of the XO-4 transiting planetary system, conducted with
the FLWO 1.2m telescope and the 8.2m Subaru Telescope.
Based on the new light curves, the refined transit ephemeris of XO-4b
is $P = 4.1250828 \pm 0.0000040$ days
and $T_c$~[BJD$_{\rm TDB}$]~$= 2454485.93323 \pm 0.00039$.
We measured the Rossiter-McLaughlin effect of XO-4b
and estimated the sky-projected angle between the stellar spin axis
and the planetary orbital axis to be
$\lambda = -46.7^{\circ} \,^{+8.1^{\circ}}_{-6.1^{\circ}}$.
This measurement of $\lambda$ is less robust than in some
other cases because the impact parameter of the transit is small,
causing a strong degeneracy between $\lambda$ and the projected
stellar rotational velocity.
Nevertheless, our finding of a
spin-orbit misalignment suggests that the migration process for XO-4b
involved few-body dynamics rather than interaction with a gaseous disk.
In addition, our result conforms with the pattern
reported by Winn et al. (2010, ApJL, 718, L145) that
high obliquities are preferentially found for stars with
effective temperatures hotter than 6250~K.
\end{abstract}
\footnotetext[*]{Based on data collected at Subaru Telescope,
which is operated by the National Astronomical Observatory of Japan.}

\section{Introduction}

Observations of the Rossiter-McLaughlin effect
(hereafter the RM effect: \cite{1924ApJ....60...15R,
1924ApJ....60...22M, 2005ApJ...622.1118O, 2007ApJ...655..550G,
2010ApJ...709..458H}) for transiting planetary systems
enable us to measure the sky-projected angle
between the stellar spin axis and the planetary orbital axis
(also called the projected spin-orbit angle, or the projected obliquity).
Previous measurements of this angle
have revealed the existence of 
well-aligned prograde orbits
($|\lambda|<10^{\circ}$;
see, e.g., HD~209458b: \cite{2000A&A...359L..13Q, 2005ApJ...631.1215W},
HD~189733b: \cite{2006ApJ...653L..69W}),
highly-inclined prograde orbits
($10^{\circ}<|\lambda|<90^{\circ}$;
XO-3b: \cite{2008A&A...488..763H, 2009ApJ...700..302W},
HD80606b: \cite{2009A&A...498L...5M, 2009ApJ...703.2091W, 2010A&A...516A..95H},
WASP-14b: \cite{2009PASP..121.1104J}, CoRoT-1b: \cite{2010MNRAS.402L...1P}),
and retrograde orbits
($90^{\circ}<|\lambda|<180^{\circ}$;
HAT-P-7b: \cite{2009PASJ...61L..35N, 2009ApJ...703L..99W},
WASP-17b: \cite{2010ApJ...709..159A}; Triaud et al. 2010,
WASP-33b: \cite{2010MNRAS.tmp..882C},
WASP-8b: \cite{2010A&A...517L...1Q},
WASP-15b: Triaud et al. 2010).

The discovery of highly tilted orbits, and retrograde orbits, has
been taken as evidence in favor of ``few-body'' theories for
planetary migration, 
such as planet-planet scattering and subsequent tidal evolution
(planet-planet scattering models,
e.g., \cite{1996Sci...274..954R, 2002Icar..156..570M, 
2008ApJ...678..498N, 2008ApJ...686..580C}), or
the Kozai mechanism \citep{1962AJ.....67..591K}
caused by a distant companion and subsequent tidal evolution
(Kozai migration models,
e.g., \cite{2003ApJ...589..605W, 2005ApJ...627.1001T, 
2007ApJ...669.1298F, 2007ApJ...670..820W}).
In this way, the RM effect has already played an important role
for studies of planetary migration processes.
Further measurements of the RM effect are important
in order to constrain theoretical models
by learning the statistical distribution
of spin-orbit alignment angles, and by discriminating
planetary migration mechanisms for individual planetary systems.
For this purpose, we have continued measurements of the RM effect and
have also initiated direct imaging observations, with the Subaru
8.2m telescope, of systems with misaligned planets to search for
evidence of perturbing stellar companions
(e.g., \cite{2007PASJ...59..763N, 2010PASJ...62..779N}).

The subject of this letter is the transiting exoplanet XO-4b,
discovered by \citet{2008arXiv0805.2921M} (hereafter MC08).
The host star XO-4 is a relatively
bright ($V = 10.7$) F5 star, and XO-4b is a giant planet
($M_p = 1.72$~$M_{\rm Jup}$, $R_p = 1.34$~$R_{\rm Jup}$; MC08)
with an orbital period of $P=4.125$~days.
We used the FLWO 1.2m telescope for photometric observations and
the Subaru 8.2m telescope for radial velocity (RV) observations.
Due to the rapid
rotation of XO-4 ($V \sin I_s = 8.8\pm0.5$~km~s$^{-1}$; MC08), the
amplitude of the RM effect for XO-4b was expected to be large.
In addition, recently \citet{2010ApJ...718L.145W} reported
that hot stars ($T_{\rm eff} > 6250$~K)
with hot Jupiters tend to have high obliquities.  Since XO-4 is a hot
star ($T_{\rm eff} = 6397 \pm 70$~K; MC08) with a hot Jupiter, we may
use this system as a further test of the proposed pattern.  We report
our photometric and RV observations in section~2, and describe
our analysis procedures in section~3.  We present our results and
discussions in section~4.  Finally, we summarize the findings of this
letter in section~5.

\section{Observations}

\subsection{Light Curves}

Photometric observations were conducted with Keplercam
on the 1.2m telescope at the Fred Lawrence Whipple Observatory (FLWO)
on Mount Hopkins, Arizona.
Three partial transits, and one complete transit, were
observed with a Sloan Digital Sky Survey (SDSS) z-band filter.
The observing dates were
UT 2008 October 5 (egress), 2008 October 13 (ingress),
2009 September 8 (ingress), and 2010 February 8 (complete).
KeplerCam is equipped with a 4096$\times$4096 CCD, giving
a square field of view 23.1' on a side.
We used 2$\times$2 binning, giving a scale of 0.68'' per
binned pixel, a readout and setup time of 11 s,
and a typical readout noise of 7 e$^{-}$ per binned pixel.
The cadence was typically 2~min for the observation on
UT 2010 February 8, and about 50~s for the other observations.
Standard IRAF\footnote{The Image
  Reduction and Analysis Facility (IRAF) is distributed by the U.S.\
  National Optical Astronomy Observatories, which are operated by the
  Association of Universities for Research in Astronomy, Inc., under
  cooperative agreement with the National Science Foundation.}
procedures were applied to the observed frames
for bias subtraction, flat-field division, and aperture photometry.
The flux of XO-4 was divided by the sum of the
fluxes of 11 nearby reference stars in the field of view.
To account for systematic effects due to differential airmass
extinction and imperfect flat-fielding, we corrected the baseline
of the observed flux as a function of time and airmass.
Photometric uncertainties were temporarily set equal to
the standard deviation of the out-of-transit relative flux on that night.
As explained in subsection 3.1, we subsequently rescaled these uncertainties to account
for time-correlated noise.
The upper panel of figure 1 shows the composite transit light curves.

\subsection{Radial Velocities}

We observed XO-4 on UT 2010 January 14, 2010 January 15,
2010 February 4, and 2010 May 17 with the High Dispersion
Spectrograph (HDS: \cite{2002PASJ...54..855N}) on
the 8.2m Subaru Telescope at Mauna Kea, Hawaii.
The observations on 2010 January 15
covered a full transit of XO-4b.
We employed the standard I2a setup of the HDS,
covering the wavelength range 4940~\AA\ $< \lambda <$ 6180~\AA\,
and a slit width of $0\farcs4$, corresponding to
a spectral resolution of about 90000.
We used the iodine gas absorption cell for
precise differential RV measurements.
The exposure times were 15--20 min, yielding a typical
signal-to-noise ratio (SNR) of approximately 80--100 per pixel.
We processed the observed frames with standard IRAF
procedures and extracted one-dimensional spectra.
We computed relative RVs following
the algorithm of \citet{1996PASP..108..500B} and
\citet{2002PASJ...54..873S}.
We estimated the internal error of each RV
based on the scatter among the RV solutions
from $\sim$4~\AA~segments of each spectrum.
The typical internal errors were 9--15~m~s$^{-1}$.
As was the case for TrES-4
\citep{2010PASJ...62..653N}, the internal errors are larger than
some other cases of spectra with
a similar SNR due to the star's broad absorption features.
The observed RVs and internal errors are given in table~1.
We note that we adopt BJD$_{\rm TDB}$, the Barycentric Julian Date in
the Barycentric Dynamical Time standard, for all time stamps
in this letter, as advocated by \citet{2010arXiv1005.4415E}.

\section{Analyses}

\subsection{Refined Transit Ephemeris}

First we fitted the FLWO transit light curves
in order to determine appropriate data weights for our
subsequent analysis, and to refine the transit ephemeris of XO-4b.
We used the analytic formula for transit light curves given
by \citet{2009ApJ...690....1O}.
The free parameters were the midtransit time $T_c$ for
each transit,
the ratio of radii of the planet and star $R_p/R_s$,
the orbital inclination $i$, and the orbital distance in units of
the stellar radius $a / R_s$.
We here adopted the orbital period $P = 4.12502$ days,
the origin of mid-transit time of XO-4b
$T_c (0) = 2454485.93295 \pm 0.00040$ in BJD$_{\rm TDB}$,
the stellar mass $M_s = 1.32$~$M_{\odot}$ reported by MC08.
We held fixed one of quadratic limb-darkening coefficients $u_1$ at
the value 0.13,
based on tables of \citet{2004A&A...428.1001C},
and allowed $u_2$ to be a free parameter.

We determined the optimal parameter values by minimizing the $\chi^2$
statistic,
\begin{eqnarray}
\chi^2 &=& \sum_i \left[ \frac{f_{i,{\rm obs}}-f_{i,{\rm calc}}}
{\sigma_{i}} \right]^2,
\end{eqnarray}
using the AMOEBA algorithm \citep{1992nrca.book.....P}.  In this
equation, $f_{i, {\rm obs}}$ are the relative flux data points, and
$f_{i, {\rm calc}}$ are the values calculated based on the analytic
formulae and a particular choice of model parameters.

We then computed the residuals for each dataset,
and averaged the residuals into $M$ bins of $N$ points.
We calculated the standard deviation of the binned data,
$\sigma_{{\rm N,obs}}$ and
\begin{eqnarray}
\sigma_{{\rm N,ideal}} = \frac{\sigma_1}{\sqrt{N}} \sqrt{\frac{M}{M-1}},
\end{eqnarray}
where $\sigma_1$ is the standard deviation of the residuals
for each dataset.
To account for increased uncertainties due to time-correlated noise
(so-called red noise; see e.g.,
\cite{2006A&A...459..249G, 2008ApJ...683.1076W}),
we computed a red noise factor
$\beta = \sigma_{{\rm N,obs}}/\sigma_{{\rm N,ideal}}$
for various $N$ (corresponding to 10-20 min), and multiplied the photometric
uncertainties of each dataset by the maximum value of $\beta$.
We estimated 1$\sigma$ uncertainties of free parameters based on
the criterion $\Delta \chi^2 = 1.0$.
Table~2 summarizes the rms, the red noise factor, and the mid-transit time
of each transit light curve.
One important result is that $i=90^{\circ}$ is allowed
within the 1$\sigma$ level.
This was problematic for subsequent RM modeling, because the two key
parameters describing the RM effect ($\lambda$ and $V \sin I_s$, see below)
are strongly degenerate when $i = 90^{\circ}$.

We then fitted all the observed mid-transit times with a linear
function
\begin{equation}
T_c (E) = E P + T_c (0),
\end{equation}
where $E$ is an integer, in order to refine the orbital period of
XO-4b.  We note that we added 0.00075 to the value of $T_c (0)$
reported by MC08 to convert the time standard from HJD$_{\rm UTC}$
(the Heliocentric Julian Date in the Coordinated Universal Time) to
BJD$_{\rm TDB}$ \citep{2010arXiv1005.4415E}.

The refined transit ephemeris for XO-4b is $P = 4.1250828 \pm
0.0000040$ days and $T_c (0) = 2454485.93323 \pm 0.00039$ in the
BJD$_{\rm TDB}$ system.  We note that the linear fit gave $\chi^2 =
6.99$ and all O-C residuals were within 1.5$\sigma$ uncertainties of
$T_c (E)$.  The uncertainty in the predicted time of the transit of UT
2010 January 15 (the night of our RM measurement) is about 1~min,
which is small enough to be of no concern for our RM modeling.

\subsection{RM Model and Joint Fitting}

To describe the RM effect of XO-4b, we derived a formula based on the procedure described in \citet{2005ApJ...631.1215W},
\citet{2009PASJ...61..991N}, and \citet{2010ApJ...709..458H}.
We assumed a rotational broadening kernel of $V \sin I_s =
8.8$~km~s$^{-1}$ (MC08) and the quadratic limb-darkening parameters
$u_1 = 0.35$ and $u_2 = 0.35$ for the iodine absorption band
based on the tables of \citet{2004A&A...428.1001C}.
The derived formula was
\begin{equation}
\Delta v = - f v_p \left[1.6159 - 0.83778
\left( \frac{v_p}{V \sin I_s} \right)^2 \right].
\end{equation}
In this formula, $\Delta v$ is the anomalous radial velocity due to
the RM effect, $f$ is the loss of light,
and $v_p$ is the subplanet velocity (the radial component of the stellar
rotation velocity
at the position being hidden by the planet).

We simultaneously fitted the Subaru RVs
and the FLWO light curves.
We fixed $P$ and $T_c (0)$ at the central values reported above.
We did not fit the RVs reported by MC08
because the errors of the MC08 RVs
are much larger than those of the Subaru RVs.
Thus, those data have little impact on the results.
In addition to the free parameters for the photometric data,
we added free parameters for
the radial velocity semiamplitude $K$,
the sky-projected angle between the stellar spin axis and
the planetary orbital axis $\lambda$,
the sky-projected stellar rotational velocity $V \sin I_s$,
and the overall velocity offset of the Subaru dataset $v_1$.
Note that we fixed the eccentricity $e$ to zero,
and the argument of periastron $\omega$ was not considered.
(When $e$ and $\omega$ were included as free parameters,
the eccentricity was found to be
consistent with zero within $1\sigma$.)
The $\chi^2$ statistic for the joint fit was
\begin{eqnarray*}
\chi^2 = \sum_i \left[ \frac{f_{i,{\rm obs}}-f_{i,{\rm calc}}}
{\sigma_{i}} \right]^2
&+& \sum_j \left[ \frac{v_{j,{\rm obs}}-v_{j,{\rm calc}}}
{\sigma_{j}} \right]^2 
+ \left[ \frac{V \sin I_s - 8.8}{0.5} \right]^2,
\end{eqnarray*}
where $v_{j, {\rm obs}}$ were the observed RVs and $v_{j, {\rm calc}}$
were RVs calculated based on the Keplerian motion and on the RM
formula given above.  The last term was an \textit{a priori} constraint
on $V \sin I_s$ to require consistency with the value reported by MC08
within its uncertainty.  This term was necessary to obtain a meaningful
conclusion about $\lambda$, as a consequence of the low impact
parameter (near-90$^\circ$ inclination) of the transit (see
subsection~3.1.).

\section{Results and Discussion}

The lower panel of figure~1 shows the RVs around the transit phase,
along with the best-fitting model (the solid line).  For reference, we
have also plotted a model RV curve under the assumption of perfect
spin-orbit alignment ($\lambda=0^{\circ}$, the dotted line).  We found
small asymmetry of the RM anomaly with respect to the midtransit time,
suggesting that the transit is not exactly equatorial and that there
is a significant spin-orbit misalignment.

Figure~2 plots the full RV dataset in BJD$_{\rm TDB}$ (left panel)
and in phase (right panel).
No significant long term trend was observed in the residuals.
Table~3 summarizes the optimal values and uncertainties
for the system parameters.
As shown in table~3, the free parameters other than
the RM-related parameters ($\lambda$ and $V \sin I_s$) are
consistent with those reported previously.
For the RM-related parameters, we found
$\lambda = -46.7^{\circ} \,^{+8.1^{\circ}}_{-6.1^{\circ}}$ and
$V \sin I_s = 8.9 \pm 0.5$~km~s$^{-1}$.
We also computed the $3\sigma$ confidence level of $\lambda$
by $\Delta \chi^2 = 9.0$ and found
$\lambda = -46.7^{\circ} \,^{+41.7^{\circ}}_{-14.8^{\circ}}$.
Figure~3 shows a $\chi^2$ contour map in $\lambda$-$V \sin I_s$ space.
The solid lines represent contours for $\Delta \chi^2 = 1.0$,
$\Delta \chi^2 = 4.0$, and $\Delta \chi^2 = 9.0$
from the inside to the outside.
The orbital axis of XO-4b appears to be tilted
relative to the rotational axis of XO-4 in sky projection
at about the $3\sigma$ level.

It should be noted that the confidence with which $\lambda$ is
found to be nonzero depends critically on the \textit{a priori}
constraint that we applied on $V \sin I_s$.  This is because when the
transit impact parameter is small (or equivalently when the orbital
inclination is near 90$^\circ$), there is a strong degeneracy between
$V \sin I_s$ and $\lambda$.  In such cases, the RV data alone are only
able to provide good constraints on the quantity $V \sin I_s
\cos\lambda$, while $V \sin I_s \sin\lambda$ is poorly constrained.
When we tried the joint fitting without
the \textit{a priori} constraint, we found
$V\sin I_s \cos\lambda = 8.4\pm 1.1$~km~s$^{-1}$,
$V \sin I_s = 10.0 \,^{+72.0}_{-2.2}$~km~s$^{-1}$,
$\lambda = -54.6^{\circ} \,^{+19.2^{\circ}}_{-35.4^{\circ}}$,
and $i = 89.1^{\circ} \pm 0.9^{\circ}$ at the 1$\sigma$ level.
In this case the confidence in the spin-orbit misalignment
was lowered, and values of $V \sin I_s$ differing strongly
from that reported by MC08 were allowed within $1\sigma$.

As long as the spectroscopic determination of $V\sin I_S$ is valid,
our result of a strong spin-orbit misalignment is secure.  However, in
order to make the result more robust, and less dependent on the
spectroscopic determination of $V\sin I_s$, it would be necessary to
obtain a more stringent constraint on the inclination $i$.
The most straightforward
way to do so would be to obtain a more precise light curve. It is
difficult to observe complete transits from the ground, because of the
comparatively long orbital period and long duration of the transits.
However, the system is in the continuous viewing zone of the Hubble
Space Telescope ($HST$), as noted by MC08. A complete transit light
curve based on $HST$ observations is therefore the highest priority
for pinning down the spin-orbit angle.

From a theoretical point of view, it is interesting that the planetary
orbit is highly tilted, given that the orbital eccentricity of XO-4b
is consistent with zero at this point in time.
However, the current bounds on the orbital
eccentricity are coarse; improving the determination of
this parameter is also a priority for future work.

Recently, \citet{2010ApJ...718L.145W} pointed out that hot Jupiters
around relatively hot ($T_{\rm eff} > 6250$~K) stars preferentially
have tilted orbits.  According to the empirical rule, XO-4, which has
an effective temperature of $T_{\rm eff} = 6397 \pm 70$~K (MC08), would
have a tilted orbit.  Our results agree with this prediction.

In addition, \citet{2010ApJ...719..602S} has recently sought evidence
for spin-orbit misalignments along the line of sight, based on the
comparison of the measured $V \sin I_S$ and the plausible range of
rotation velocities for a star of the given mass, radius, and age.  He
argued that if a value $\Theta \equiv (V \sin I_{s,sim}-V \sin
I_{s,obs})/\sqrt{\sigma^2_{sim} + \sigma^2_{obs}}$ is larger than 2.9
(a threshold based on the SPOCS catalog: \cite{2005ApJS..159..141V}),
the stellar rotational axis is very likely to be inclined with respect
to the line of sight.  For a transiting planet host, this would imply
a spin-orbit misalignment.  (He also found that hot stars were more
likely to have high obliquities, although he phrased the correlation
in terms of stellar mass rather than effective temperature.) For XO-4,
the value is $\Theta = 2.61$, which is only just below the (somewhat
arbitrary) threshold of 2.9.  This suggests that XO-4 has a smaller
projected rotational velocity than expected, and can be taken as
supporting evidence for a spin-orbit misalignment in the XO-4 system.

\section{Summary}

We conducted photometric transit observations with the FLWO 1.2m
telescope, and RV observations (including an RM measurement) with the
8.2m Subaru Telescope, for the exoplanet XO-4b.  We refined
the transit ephemeris for XO-4b based on the new transit light curves,
and modeled the RM effect of XO-4b by jointly fitting
the photometry and RVs.
In conjunction with the previous spectroscopic estimate of
the projected stellar rotational velocity $V \sin I_s$, we found
evidence for spin-orbit misalignment in this system.  In this case the
RM data do not provide an strong independent constraint on $V \sin
I_s$, because of the low impact parameter of the transit. To
complement the RM measurement in this system, a more stringent
constraint on the orbital inclination would be useful to break the
fitting degeneracy between $\lambda$ and $V \sin I_s$.

The spin-orbit misalignment of XO-4b suggests that this planet
migrated through a mechanism that excited its orbital inclination,
although the eccentricity of XO-4b seemed to be quite low.
Furthermore, the spin-orbit misalignment of XO-4b is consistent with
the findings by \citet{2010ApJ...718L.145W} and
\citet{2010ApJ...719..602S} that hot stars with hot Jupiters have high
obliquities.  Further follow-up observations to determine a more
accurate spin-orbit alignment angle in this system would be still
interesting.

This letter is based on data collected at Subaru Telescope,
which is operated by the National Astronomical Observatory of Japan.
We acknowledge the support for our Subaru HDS observations
by Akito Tajitsu, a support scientist for the Subaru HDS.
The data analysis was in part carried out on common use data analysis
computer system at the Astronomy Data Center, ADC,
of the National Astronomical Observatory of Japan.
N.N. and T.H. are supported by a Japan Society for Promotion of Science
(JSPS) Fellowship for Research (PD: 20-8141, DC1: 22-5935).
R.S. is funded by Caja de Ahorros y Pensiones de Barcelona,
"la Caixa", under the Fellowship Program to extend graduate studies
in the United States.
We gratefully acknowledge support from the NASA Origins program through
award NNX09AD36G (to J.N.W.) and NNX09AB33G (to M.J.H. and J.N.W.),
as well as the MIT Class of 1942.
KeplerCam was developed with partial support from the Kepler mission
under NASA Cooperative Agreement NCC2-1390 (PI: D. Latham).
M.T. is supported by the Ministry of Education, Science,
Sports and Culture, Grant-in-Aid for
Specially Promoted Research, 22000005.
We wish to acknowledge the very significant cultural role
and reverence that the summit of Mauna Kea has always had within
the indigenous people in Hawai'i.



\begin{figure}[thb]
 \begin{center}
  \FigureFile(120mm,120mm){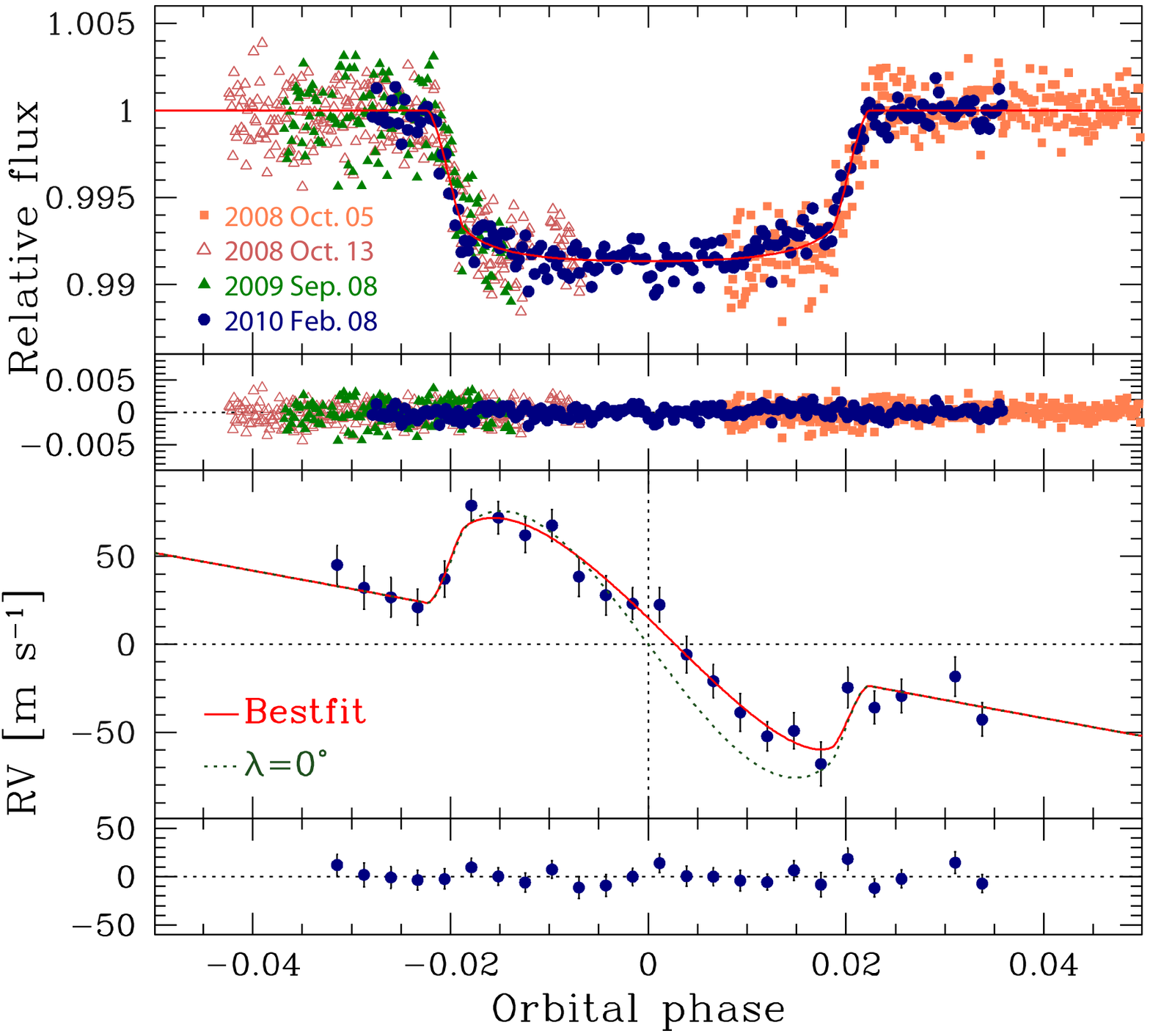}
 \end{center}
  \caption{
  Top panel: Phased transit light curves taken with the FLWO 1.2m
  telescope. Four different epoch data are combined.
  The solid line represents the best-fit curve
  based on the optimal free parameters listed in table~3.
  Second panel: Residuals of the photometric data from the best-fit model.
  Third panel: Subaru radial velocities and the best-fit curve
  (the solid line) as well as a model curve which assumed
  $\lambda=0^{\circ}$ (the dotted line).
  Bottom panel: Residuals of radial velocities from the best-fit model.
  }
 \begin{center}
  \FigureFile(120mm,120mm){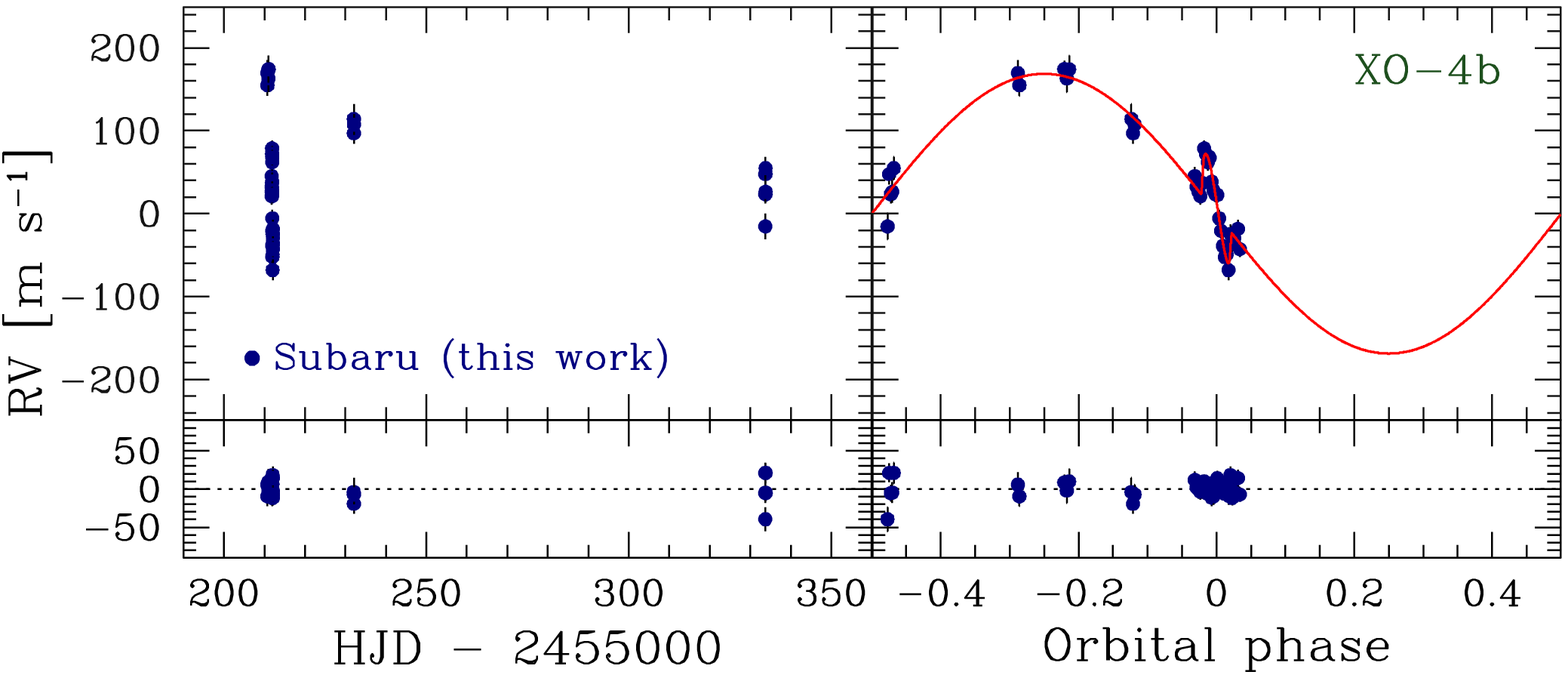}
 \end{center}
  \caption{
  Upper panels: Subaru radial velocities in BJD$_{\rm TDB}$ (left panel)
  and in phase (right panel).
  Lower panels: Residuals from the best-fit model.
  No significant long term trend is apparent.
  }
\end{figure}

\begin{figure}[thb]
 \begin{center}
  \FigureFile(100mm,100mm){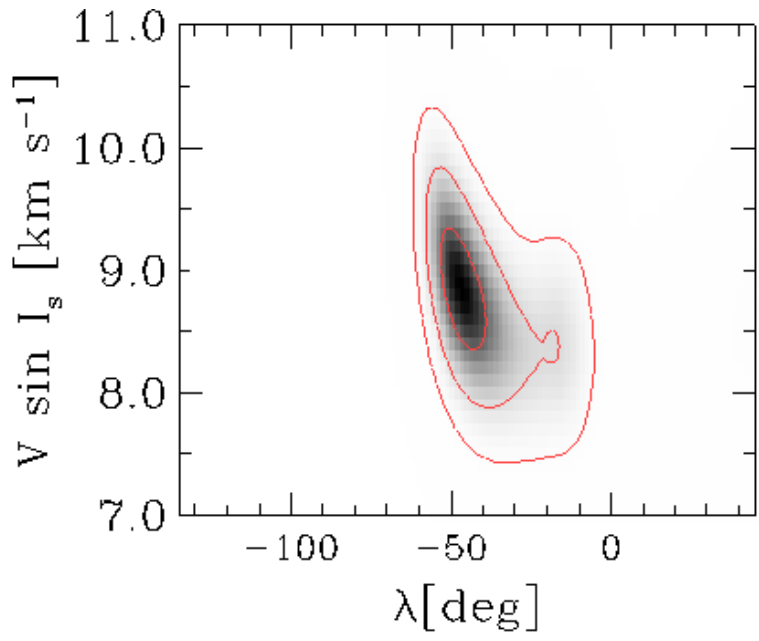}
 \end{center}
  \caption{
  A $\chi^2$ contour map in $\lambda$-$V \sin I_s$ space.
  The solid lines represent contours for $\Delta \chi^2 = 1.0$,
  $\Delta \chi^2 = 4.0$, and $\Delta \chi^2 = 9.0$
  from the inside to the outside.
  }
\end{figure}

\begin{table}[htb]
\caption{Radial velocities obtained with the Subaru/HDS.}
\begin{center}
\begin{tabular}{lcc}
\hline
Time [BJD$_{\rm TDB}$]  & Value [m~s$^{-1}$] & Error [m~s$^{-1}$]\\
\hline
2455210.75952 &  169.41 &  16.13 \\
2455210.76897 &  154.86 &  12.81 \\
2455211.03700 &  174.33 &  10.06 \\
2455211.05170 &  163.09 &  16.51 \\
2455211.06638 &  173.97 &  16.29 \\
2455211.81802 &   45.11 &  10.91 \\
2455211.82923 &   32.15 &  12.19 \\
2455211.84044 &   26.67 &  11.31 \\
2455211.85165 &   21.03 &  10.25 \\
2455211.86285 &   37.17 &  10.35 \\
2455211.87405 &   78.81 &   9.42 \\
2455211.88526 &   72.01 &   9.11 \\
2455211.89646 &   61.92 &   9.83 \\
2455211.90768 &   67.59 &   9.19 \\
2455211.91890 &   38.45 &  11.15 \\
2455211.93011 &   27.91 &  11.16 \\
2455211.94132 &   23.14 &   9.01 \\
2455211.95255 &   22.51 &   9.83 \\
2455211.96377 &   -5.80 &  10.34 \\
2455211.97497 &  -20.92 &   9.45 \\
2455211.98619 &  -38.70 &  10.70 \\
2455211.99740 &  -52.24 &   8.29 \\
2455212.00861 &  -49.20 &  10.16 \\
2455212.01982 &  -68.05 &  12.48 \\
2455212.03102 &  -24.63 &  11.57 \\
2455212.04223 &  -35.98 &   9.33 \\
2455212.05343 &  -29.42 &   9.41 \\
2455212.07586 &  -18.34 &  11.22 \\
2455212.08708 &  -42.78 &   9.52 \\
2455232.06436 &  114.02 &  18.42 \\
2455232.07383 &   97.09 &  13.09 \\
2455232.08330 &  107.35 &  12.66 \\
2455333.73171 &  -15.49 &  15.33 \\
2455333.74176 &   47.77 &  12.39 \\
2455333.75088 &   23.27 &  11.23 \\
2455333.76001 &   26.43 &  13.70 \\
2455333.76915 &   54.67 &  13.64 \\
\hline
\end{tabular}
\end{center}
\end{table}

\begin{table*}[tht]
\caption{Summary of the XO-4 transit light curves.}
\begin{center}
\begin{tabular}{cccccc}
\hline
UT Date & Epoch & RMS & Red noise factor & Mid-transit time & Note \\
 & $E$ & $10^4 \sigma$ & $\beta$ & BJD$_{\rm TDB}$ &  \\
\hline
2008 Jan. 20 & 0 &  --- & --- & $2454485.93295 \pm 0.00040$ & MC08 \\
2008 Oct. 05 & 63  & 12.7 & 1.293 & $2454745.81494 \pm 0.00105$ & Egress \\
2008 Oct. 13 & 64  & 15.4 & 1.497 & $2454749.93969 \pm 0.00114$ & Ingress \\
2009 Sep. 08 & 144 & 17.2 & 1.002 & $2455079.94672 \pm 0.00110$ & Ingress \\
2010 Feb. 08 & 182 &  8.2 & 1.807 & $2455236.69753 \pm 0.00065$ & Complete \\
\hline
\end{tabular}
\end{center}
\end{table*}

\begin{table*}[tht]
\caption{System parameters of XO-4 derived from the joint fitting.}
\begin{center}
\begin{tabular}{lccc}
\hline
Parameter & Value & Uncertainty & Note \\
\hline
\multicolumn{4}{l}{Adopted parameters}    \\
$P$ [days] & 4.1250828  & $\pm$0.0000040  & this work \\
$T_c (0)$ [BJD$_{\rm TDB}$] & 2454485.93323 & $\pm$0.00039 & this work \\
$M_s [M_{\odot}]$ & 1.32  & $\pm$ 0.02   & MC08   \\
$V \sin I_s$ [km~s$^{-1}$] & 8.8  & $\pm$0.5   & MC08   \\
$u_1$ & 0.13  & ---   &  \citet{2004A&A...428.1001C}  \\
$e$ & 0  & ---   & assumed \\
\hline
\multicolumn{4}{l}{Best-fit parameters with
the \textit{a priori} constraint} \\
$K$ [m s$^{-1}$] 
& 168.6 & $\pm$ 6.2  &   \\
$V \sin I_s$ [km s$^{-1}$]
& 8.9  & $\pm$0.5 &   \\
$\lambda$ [$^{\circ}$]
& -46.7  & $^{+8.1}_{-6.1}$ & $^{+41.7}_{-14.8}$ ($3\sigma$) \\
$a/R_s$
& 7.68 & $\pm$ 0.11 &   \\
$R_p/R_s$
& 0.0881 & $\pm$ 0.0007 &   \\
$i$ [$^{\circ}$]
& 88.8 & $\pm$ 0.6 &   \\
$u_2$ 
& 0.35  & $\pm$ 0.11  &   \\
$v_1$ [m s$^{-1}$] 
& -0.1  & $\pm$ 2.9  &    \\
rms [m s$^{-1}$] 
& 11.50  & ---  &    \\
$\chi^2$/$\nu$ (RV) 
& 31.63/37  & ---   &    \\
$\chi^2$/$\nu$ (LC) 
& 487.22/906  & ---   &    \\
\hline
\multicolumn{4}{l}{Derived planet parameters}  \\
$M_p [M_{Jup}]$ & 1.78  & $\pm$ 0.08  & this work   \\
$R_p [R_{Jup}]$ & 1.33  & $\pm$ 0.05  & this work   \\
\hline
\multicolumn{4}{l}{Comparison with previous literature}  \\
$K$ [m s$^{-1}$] 
& 163 & $\pm$ 16  &  MC08 \\
$a/R_s$
& 7.7 & $\pm$ 0.2 & MC08  \\
$R_p/R_s$
& 0.089 & $\pm$ 0.001 & MC08  \\
$i$ [$^{\circ}$]
& 88.7 & $\pm$ 1.1 &  MC08 \\
$M_p [M_{Jup}]$ & 1.72  & $\pm$ 0.20  & MC08   \\
$R_p [R_{Jup}]$ & 1.34  & $\pm$ 0.048  & MC08   \\
$u_2$ 
& 0.36  & --- & \citet{2004A&A...428.1001C}  \\
\hline
\end{tabular}
\end{center}
\end{table*}

\end{document}